\documentclass[a4paper]{article}

\usepackage{amssymb}
\usepackage{INTERSPEECH2018}
\usepackage[colorlinks, linkcolor=red, anchorcolor=blue, citecolor=green]{hyperref}

\title{MTGAN: Speaker Verification through Multitasking Triplet \\ Generative Adversarial Networks}
\name{Wenhao Ding and Liang He}
\address{Department of Electronic Engineering, Tsinghua University, China}
\email{dwh14@mails.tsinghua.edu.cn, heliang@mail.tsinghua.edu.cn}

\begin{document}
\maketitle

\begin{abstract}

\noindent In this paper, we propose an enhanced triplet method that improves the encoding process of embeddings by jointly utilizing generative adversarial mechanism and multitasking optimization. We extend our triplet encoder with \textit{Generative Adversarial Networks} (GANs) and softmax loss function. GAN is introduced for increasing the generality and diversity of samples, while softmax is for reinforcing features about speakers. For simplification, we term our method \textit{Multitasking Triplet Generative Adversarial Networks} (MTGAN). Experiment on short utterances demonstrates that MTGAN reduces the verification equal error rate (EER) by \textit{67\%} (relatively) and \textit{32\%} (relatively) over conventional i-vector method and state-of-the-art triplet loss method respectively. This effectively indicates that MTGAN outperforms triplet methods in the aspect of expressing the high-level feature of speaker information.
\end{abstract}
\noindent\textbf{Index Terms}: generative adversarial networks, speaker verification, triplet loss

\section{Introduction}

Automatic Speaker Verification (ASV) refers to the process of identifying the speaker's ID of an unknown utterance given a registered voice database. As an important non-contact biometric identification technique, it has been widely studied \cite{ubm, ivector, softmax, triplet}.

In the past few years, the field of ASV has formed a mainstream with i-vector/PLDA \cite{ivector, plda}. However, plenty of works have found that end-to-end systems composed of Deep Neural Networks (DNN) surpass traditional methods in some aspects, especially under short-utterance condition. Besides, speaker verification of short voice is of great practical value, which motivates us to research on DNN methods. 

Recently, some metric learning methods with DNNs attract a lot of attention. Triplet loss is one of them and is popular in the field of pattern recognition because of FaceNet \cite{facenet}, which is a novel method of face recognition. After that, Zhang \emph{et al.} \cite{triplet} apply this method to speaker verification. Triplet method has been proved to be useful and large amount of works \cite{improve1, improve2, improve3} are improved on the basis of it.

The essential thought of triplet loss is to minimize intra-class distance while maximizing inter-class distance. Theoretically, it is effective for all classification tasks, but considering limited training samples, reverberation and ambient noise during recording, triplet loss has limitations on the task of speaker verification. In the absence of any guidance or restriction, encoders with vanilla triplet loss usually extract features unrelated to the speaker's ID, resulting in poor performance. Furthermore, generalization ability is important for zero-shot learning. Training encoders entirely on training set without any augmentation makes triplet methods less general on a test set. To address aforementioned issues, we propose to enhance triplet loss with multitasking learning and generative adversarial mechanism.

\begin{figure*}[]
\centering
\includegraphics[width=16cm]{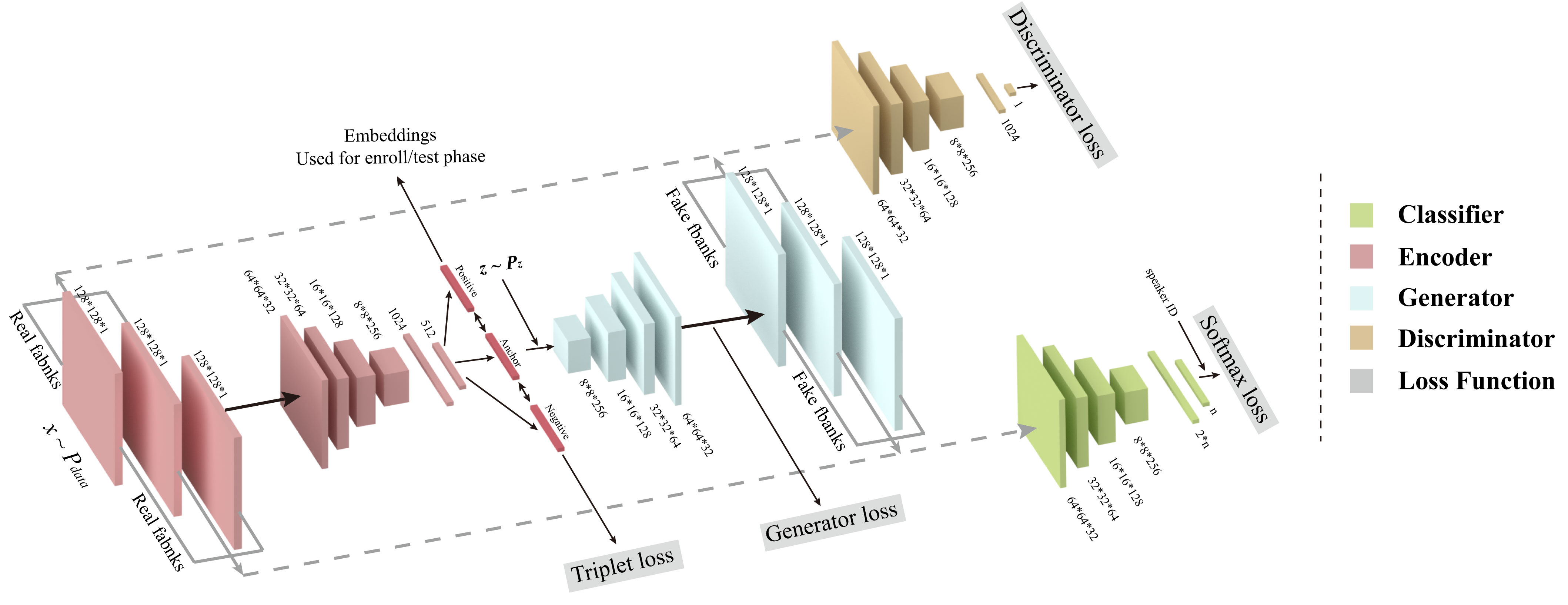}
\caption{\textbf{Architecture of Multitasking Triplet Generative Adversarial Networks.} Parameters are shared among the networks for triplet samples. In enroll/test phase, embeddings produced by the encoder are used to calculate the distance. Best viewed in color.}
\label{fig:CNN}
\end{figure*}

As for our architecture (shown in Figure~\ref{fig:CNN}), two more modules are introduced in addition to the basic encoder. First, we tail a \textit{Conditional GAN} behind the encoder. The generator produces new samples from embeddings of the encoders and random noise. Merging encoders with GAN is similar to the framework of \cite{vaegan1} and \cite{vaegan2}, which have proved their superiority. After passing through an encoder-decoder structure with noise, new samples have more generalization ability and variety in terms of speech context and unrelated environment information. The discriminator guarantees the authenticity and similarity of the generated samples, while the features of speaker remain because of the following restrictions. A classifier takes samples both from the generator and raw data as input. The last layer of this classifier is used for softmax loss, whose labels are speakers' ID of the training set. Such a module improves the ability of the encoder to extract distinctive features of speakers.

We train and test our method on two different datasets to analyse the transferability of the algorithm. Our baselines include a i-vector/PLDA system, a softmax method \cite{softmax} and a triplet method \cite{triplet}. Experimental results show that our algorithm achieves \textit{1.81\%} of EER and \textit{92.65\%} of accuracy that are much better than baseline systems. Through more extensive experiments (see experiments section), we confirms that MTGAN has more ability to extract speaker related features than vanilla triplet loss methods.

\section{Related Works}

\subsection{Deep Neutral Networks}

The appearance of \textit{d-vector} \cite{softmax} signifies the birth of ASV systems under the entire DNN framework. It is a milestone in the field of ASV which leads a large amount of works about DNN. After that, more and more works \cite{embeddings, deepspeaker, wangdong} achieve as good results as i-vector/PLDA methods. For instance, \cite{wangdong} presents a convolutional time-delay deep neural network structure (CT-DNN) and claims they are much better than i-vector systems in the case of short time speech.

In this area, a lot of works focus on the adjustment of the network structure and the utilization of new training technologies. However, in terms of zero-shot tasks like ASV, whose training set and test set are irrelevant, more suitable method should be proposed rather than optimizing the network structure. \cite{deepspeaker} claims that only using softmax loss like \cite{softmax} and \cite{wangdong} leads to poor performance on test sets that are very different from their training sets.

\subsection{Triplet Metric Learning}

For the sake of tackling zero-shot problems, triplet loss is proposed for the first time in \cite{original-triplet}. Although it has appeared for a long time, there are still many subsequent works \cite{improve1, improve2, improve3}. \cite{improve1} adopted a multi-channel approach to enhance the tightness of intra-class samples. \cite{improve2} proposed a structure of quadruplet network to improve the transferability of triplet loss on the test set. Some other works like \cite{improve3} directly modify the definition of distance and margin.

Inspired by FaceNet \cite{facenet}, which improved the sampling method of triplet loss, \cite{triplet} combined triplet loss with \textit{ResNet} \cite{resnet} and applied it to ASV for the first time. After that, \cite{TRISTOUNET} also proposed a structure called \textit{TRISTOUNET} for speaker verification using a combination of bidirectional LSTM and triplet loss. \cite{deepspeaker} proposed \textit{Deep Speaker} to tackle both text-dependent and text-independent tasks. Deep Speaker also proves that pre-trained softmax network is conducive to improve triplet methods.

Methods mentioned above have adopted a variety of ways of improving, but none of them combine triplet loss with other multitasking methods. Despite \textit{deepspeaker} uses pre-trained softmax network, there is only one loss item during the training process.

\subsection{Generative Adversarial Network}

GAN \cite{gan} is a framework based on game theory presented in 2014. After the proposal of original GAN, dozens of variants \cite{cgan, dcgan, infogan, wgan} appear and are widely used in many fields.

The framework of GAN contains two players, one is generator and the other is discriminator. Generator and discriminator play the following minimax game with value function \textit{V(G, D)}:
\begin{equation}
\begin{split}
	\mathop{\min}_{G}\mathop{\max}_{D}V(G,D) = {\mathbb{E}}_{x\sim{p}_{data}(x)}[logD(x)] \\
	+ {\mathbb{E}}_{z\sim{p}_{z}(z)}[log(1-D(G(z)))]
\end{split}
\end{equation}
where \textit{z} is a random noise that is introduced to avoid mode collapse. \textit{G(z)} is fake samples generated from the generator. The first item of the equation represents the probability that the discriminator holds real sample is true, and the second item represents the probability that the discriminator holds fake sample is false.

Intuitively, GANs are usually used for generative tasks, but recently there are some works using GAN for classification tasks \cite{vaegan1, vaegan2}. Our architecture is similar to \cite{vaegan1} that combines an encoder with GAN.

Most applications of GAN are related to computer vision. However, researchers utilize GAN in the field of speech lately. \cite{segan} and \cite{denoise} apply GAN to denoise and enhance voice. \cite{speech-gan} improve the process of speech recognition with GAN. Some people also combine triplet loss with GAN \cite{tripletgan1, tripletgan2} to explore new applications. Concretely, \cite{tripletgan1} proposed a triplet network to generate samples specially for triplet loss. \cite{tripletgan2} proposed \textit{TripletGAN} to minimize the distance between real data and fake data while maximizing the distance between different fake data.

In the field of speech, most previous work with GAN is about data enhancement. To the best of our knowledge, no one has proposed to enhance triplet loss with GAN for the task of speaker verification. 

\begin{figure*}[th]
\centering
\includegraphics[width=15cm]{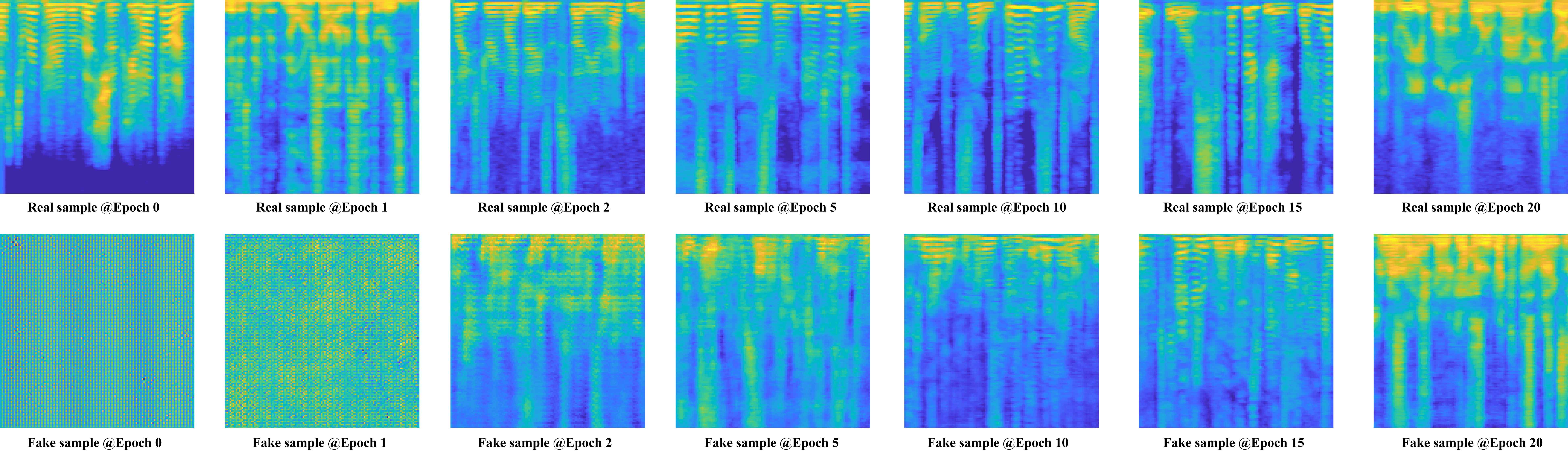}
\caption{\textbf{Generative samples during the training process.} The first row contains fbanks that fed into the encoder, and the second row contains fake samples generated from the generator. From left to right in the order of training.}
\label{fig:GAN}
\end{figure*}

\section{Multitasking Triplet Generative Adversarial Network}

\subsection{Network Architecture}

Figure~\ref{fig:CNN} displays the architecture of our network. It consists of four modules and all of them have already been marked as different colors.
\begin{itemize}
\item \textit{Encoder}: This module is used to extract features from samples. The last fully-connected layer of it outputs a 512 dimension embedding, which represents speaker information of the original sample. In the enroll/test stage, this embedding is used for calculating distance between unknown utterance and registered utterances.
\item \textit{GAN}: More specifically, this is a GAN with conditional architecture. The inputs of the generator are not only the random noise but also embeddings from the encoder. The output of the generator is fake samples that are expected to look like original samples. Discriminator has two kinds of inputs, one is the real sample and the other is the fake sample from a generator. 
\item \textit{Classifier}: Similarly, we feed both fake samples and real samples into the classifier module. The output of this module is a one-hot vector, whose size is equal to the number of the speaker in training set. 
\end{itemize}

In the whole framework, we only use convolution layer and fully-connected layer. The kernel size of all convolution layer is \textit{5$\times$5} and we also utilize batch normalization. \cite{batchnorm}

\subsection{Loss Function}

The loss function of our algorithm has four components, and each of them has a weight coefficient. The first one is a standard triplet loss that has been fully explained in \cite{triplet} :
\begin{equation}
\sum_{i}^{N}[{\parallel f(x_{i}^{a})-f(x_{i}^{p})\parallel}_{2}^{2}-{\parallel f(x_{i}^{a})-f(x_{i}^{n})\parallel}_{2}^{2} + \alpha]_{+}
\end{equation}
where \textit{a} (anchor) and \textit{p} (positive) represent samples from the same class, while \textit{a} (anchor) and \textit{n} (negative) represent samples from different classes. \textit{$\alpha$} is a hyper-parameter margin, which defines the distance between intra-class  samples and inter-class samples. It is set to \textit{0.2} in our experiments. In our algorithm, we take advantage of cosine distance to measure the differences between embeddings that produced by the encoder. The second term is the softmax loss of the classifier, whose labels are speakers' ID of the training set. The sum of triplet loss and softmax loss are named as encoder loss function to measure the encoder's ability of extracting features. The last two loss functions come from the generator and discriminator of GAN, thus the whole loss function will be expressed as:
\begin{equation}
	\mathbb{L}_{MTGAN}=\underbrace{\omega_{1}\mathbb{L}_{Triplet}+\omega_{2}\mathbb{L}_{Softmax}}_{Encoder\ Loss}+\omega_{3}\mathbb{L}_{G}+\omega_{4}\mathbb{L}_{D}
\end{equation}
where \textit{$\omega_{i}$} represents the weight coefficient of each item. In consideration of the generating diversity, we optimize the generator more times than discriminator. All of the \textit{$\omega_{i}$} are determined through experiments, and we set them to \textit{0.1, 0.2, 0.2, 0.5} respectively.

\subsection{Triplet Sampling Method}

The accuracy and convergence speed of triplet approach heavily depend on sampling method, and this problem has been detailedly discussed in \cite{sampling-matter}. There are tremendous combinations between all utterances totally, as a result, it is impossible to consider all possibility. \cite{facenet} proposed to use semi-hard negative exploration to sample triplet pairs, and \cite{triplet} followed it. This method searches triplet pairs inside one mini-batch, thus it is effective and timesaving. \textit{Deep Speaker} \cite{deepspeaker} also propose to search anchor-negative pairs across multiple GPUs.

After comparing random selection with semi-hard negative selection \cite{facenet} (details in experiment section), we find that the selecting method does not matter as long as large amounts of people are used in one epoch. Thus, we directly use random sampling method in our algorithm. Totally, we obtain \textit{n*A*P*K*J} triplet pairs in one epoch, where \textit{n} represents the number of people selected, \textit{A} is the number of \textit{Anchor}, \textit{P} is the number of \textit{Positive}, \textit{K} is the number of other classes inside \textit{n} and \textit{N} is the number of \textit{Negative} of each \textit{K}.

\subsection{Details about Training Networks}

In the stage of preprocessing, we extract \textit{mel-fbank} from raw audio slice. The length of each slice is \textit{2s} and we use \textit{128} mel-filters, thus the dimension of the input is \textit{128$\times$128}. Admittedly, GAN is difficult to train because of its instability, especially in our multitasking situation. Like most works, we choose to modify the DCGAN architecture proposed by \cite{dcgan}, and utilize state-of-the-art training skills of WGAN-GP \cite{wgan}. Some generative samples during the training process are shown in Figure~\ref{fig:GAN}.

\section{Experiments and Discussion}

\subsection{Datasets and Baselines}

The dataset that we utilize for training is \textit{Librispeech} \cite{librispeech}, which consists of "clean" part and "other" part. We use "other" part only for the experiment that explores the influence of the number of speakers. The test dataset is \textit{TIMIT} \cite{timit}, because this dataset covers all English phonemes. The reason we train and test on different datasets is to explore the transferability of algorithms. In terms of the evaluation settings, we randomly choose \textit{3} utterances for enrollment and \textit{7} utterances for test.
\begin{table}[th]
\caption{EER (\%) and accuracy (\%) of different systems }
\label{tab:exp1}
\centering
\begin{tabular}{ccc}
	\toprule
	\textbf{Methods} & \textbf{Equal Error Rate} & \textbf{Accuracy}  \\
    	\midrule
    	i-vector/Cosine                       &     8.48\%               &  81.92\% \\
    	i-vector/PLDA                         &     5.61\%               &  85.78\% \\
	Softmax loss \cite{softmax}    &     3.61\%               &  88.23\% \\
    	Triplet loss \cite{triplet}           &     2.68\%               &  90.45\% \\
    	MTGAN                                  &     \textbf{1.81\%}   &  \textbf{92.65\%}  \\
    	\bottomrule
\end{tabular}
\end{table}

We have four baselines for comparison. Two of them are i-vector systems, another is a supervised softmax system \cite{softmax} and the last one is a triplet system \cite{triplet}. 

\subsection{Performance Comparison Experiments}

In this section, we compared our method with baselines under the same experiment settings (training with \textit{1252} people of Librispeech), and the result is displayed in Table~\ref{tab:exp1}. We use EER and ACC as our evaluation criteria. EER evaluates the overall performance of the system and ACC reveals the best result for us. For a more comprehensive assessment, we plot detection error trade-off (DET) curves of all five systems (shown in the left part of Figure~\ref{fig:graph1}).

Through the results in Table~\ref{tab:exp1}, we summarize that triplet method \cite{triplet} indeed outperforms i-vector and softmax methods. However, our method achieves a better result than \cite{triplet} and has faster convergence speed. Through the analysis, we think the simple triplet method is limited by the ability of feature extraction and is poor in the performance of data transfer. In the later period of training, triplet loss of \cite{triplet} is close to zero (not overfitting). This phenomenon indicates that it has reached the limit of the speaker verification task with current features. The encoder extracts features not only from the speaker information but also from other independent factors.

\subsection{Ablation Experiments}

In this section, we did more ablation experiments to prove that our framework is feasible. Results under different conditions are shown in Table~\ref{tab:exp2}. First, we verify the necessity of each module in our structure. We removed three modules once at a time and carried out experiments under the same settings. Results prove that the structure after the removal of modules cannot behave as effectively as MTGAN. Among three situations, the removal of classifier has the greatest impact, which means softmax loss is important for improving feature extraction process.

\begin{figure}[t]
\centering
\includegraphics[width=8cm]{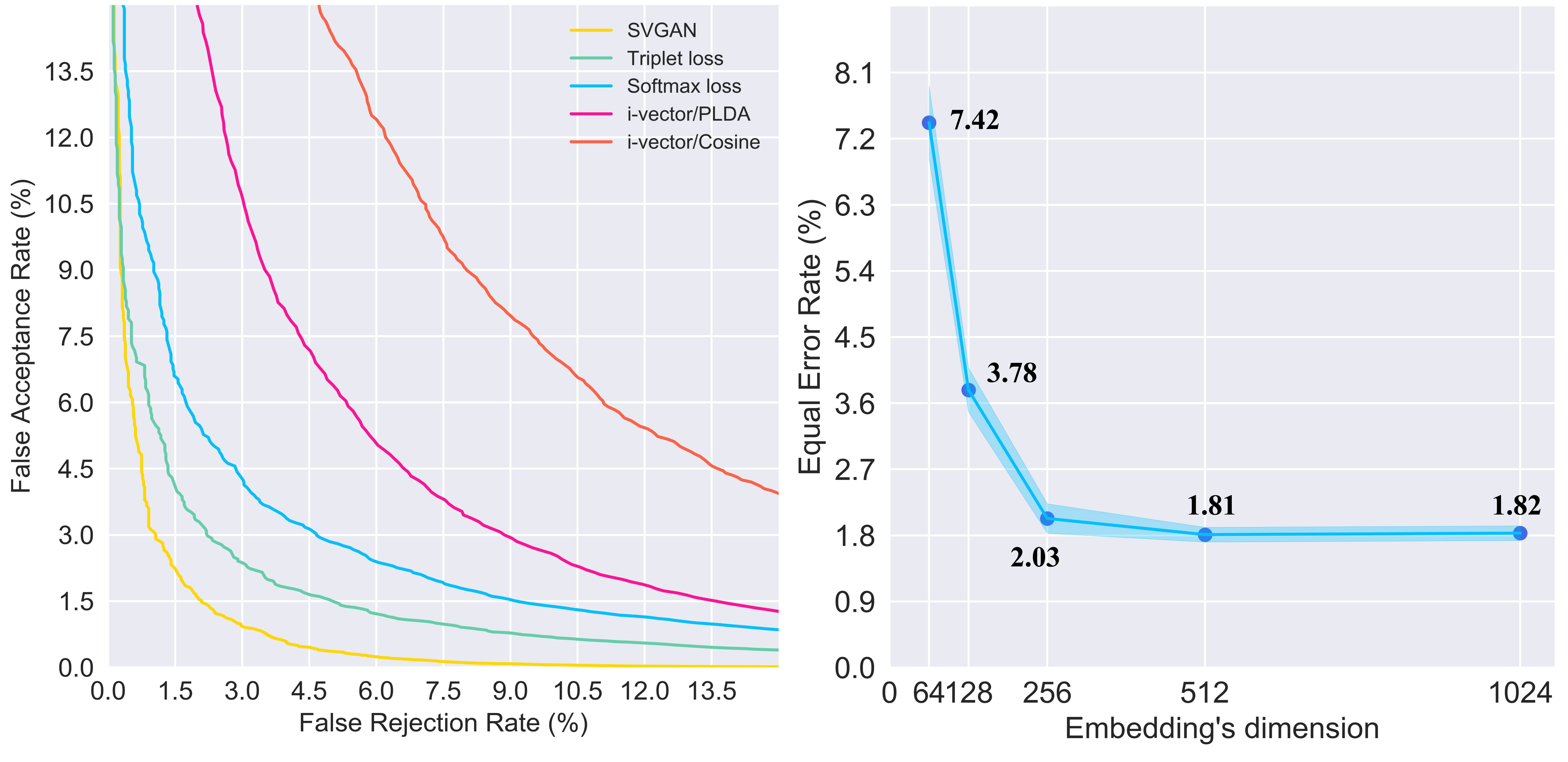}
\caption{\textbf{Left:} DET curves. Results of five methods are displayed. Two back-end methods are used for i-vector system. \\
	     \textbf{Right:} EER (\%) with different embedding dimensions. Five kinds of dimensions of the embeddings are displayed.}
\label{fig:graph1}
\end{figure}

\begin{figure}[t]
\centering
\includegraphics[width=8cm]{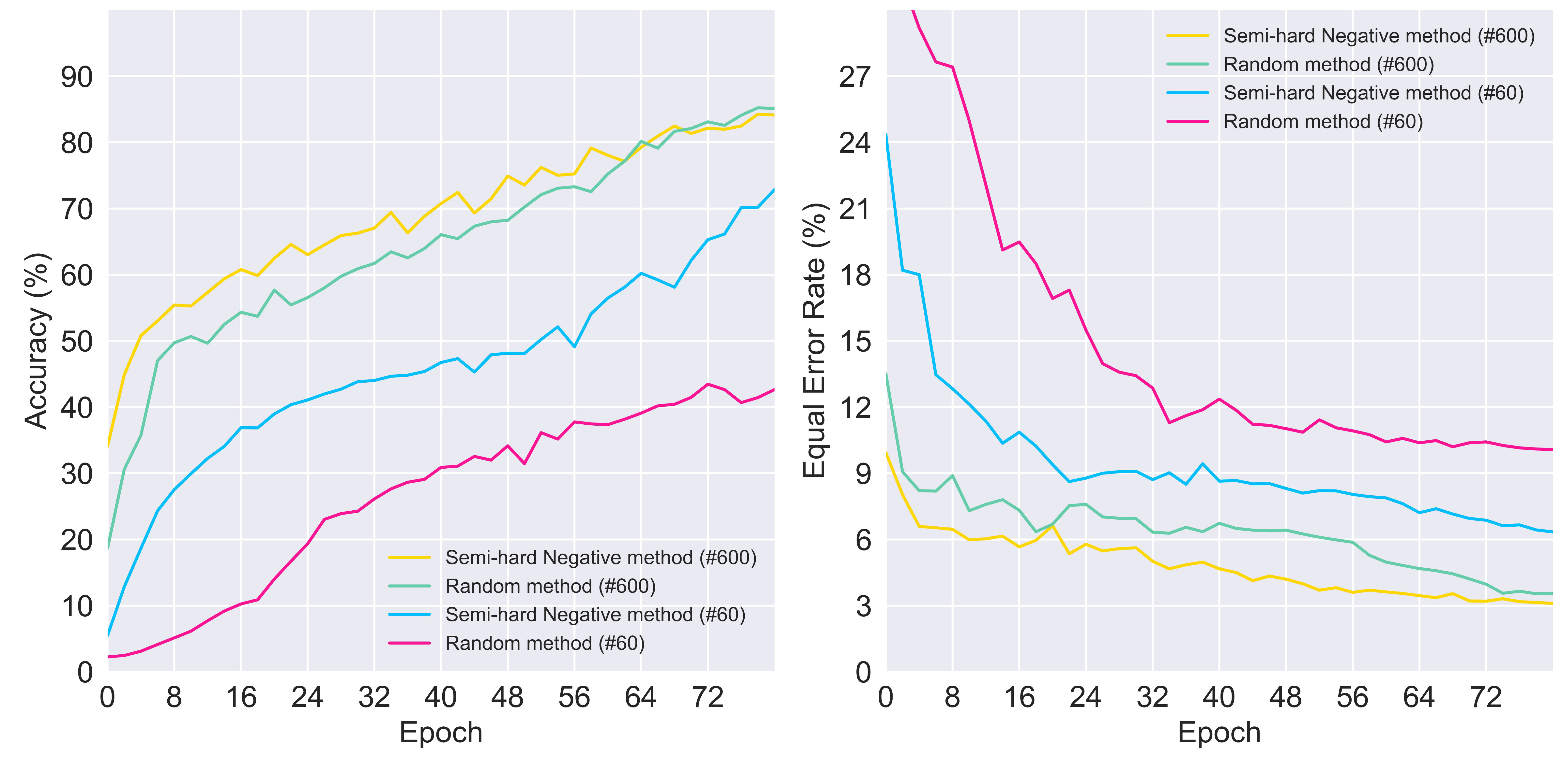}
\caption{\textbf{Changes of ACC (\%) and EER (\%) with the time of training.} We choose the first 80 epochs to show the trend.}
\label{fig:graph2}
\end{figure} 

Then we compared the difference between the random sampling method and the semi-hard negative method proposed by \cite{facenet}. The network architecture we applied was \textit{Inception-Resnet-v1}, and we tested on selecting \textit{60} and \textit{600} people of each epoch for both methods. Results reported in Table~\ref{tab:exp2} shows that the gap between the two methods is very tiny in the case of a large number of people. We also find that the number of selected people has more influence than the number of samples from the same person. EER and ACC of each epoch are displayed in Figure~\ref{fig:graph2}.

Embedding's dimension is also an important factor that influences the expressing ability of the system. Therefore, we compared the EERs of five different dimensions and the results are displayed in the right part of Figure~\ref{fig:graph1}

\begin{table}[th]
  \caption{Ablation experiments with different conditions}
  \label{tab:exp2}
  \centering
  \begin{tabular}{cccc}
    \toprule
    \textbf{Conditions} & \textbf{EER} & \textbf{ACC}  & \textbf{Convergence} \\
    \midrule
    w/o GAN                  &   2.04\%    &  90.17\%   &  60 epoch\\
    w/o softmax loss      &   3.34\%    &  88.63\%   &  80 epoch\\
    w/o triplet loss          &   2.71\%    &  89.51\%   &  60 epoch\\
    MTGAN                           &   \textbf{1.81\%}   &  \textbf{92.65\%}  & 100 epoch \\
    \midrule
    Random (\#60)               &   3.13\%    &  85.26\%   &  550 epoch\\
    Semi-hard (\#60)            &   2.90\%    &  88.73\%   &  500 epoch\\
    Random (\#600)             &   2.75\%    &  90.03\%   &  250 epoch\\
    Semi-hard (\#600)          &   \textbf{2.68\%}    &  \textbf{90.45\%}   &  200 epoch\\
    \midrule
    1252 people     &   1.81\%    &  92.65\%   &  70 epoch\\
    2484 people     &   \textbf{1.33\%}    &  \textbf{94.27\%}   &  100 epoch\\
    \bottomrule
  \end{tabular}
\end{table}

The last experiment is to explore the impact of the number of people in training set. We added the "other" part of \textit{Librispeech} to the training set (\textit{2484} in total) and did the same experiment with the one that had \textit{1252} people. Although the convergence speed became slower, EER and ACC increased after enlarging training set. We cannot fail to note a phenomenon: the output layer of classifier is related to the number of training speaker. The size of the network will increase if we use a larger dataset to train the model.

\section{Conclusion}

In this study, we present a novel end-to-end text-independent speaker verification system on short utterances, which is named MTGAN. We extend triplet loss with classifier and generative adversarial networks to form a multitasking framework. Triplet loss is designed for clustering, while GAN and softmax loss help with extracting features about speaker information.

Experimental results demonstrate that our algorithm achieves lower EER and higher accuracy over i-vector methods and triplet methods. Besides, our method has a faster convergence speed than vanilla triplet methods.Through more ablation experiments, we get other conclusions. We confirm that softmax loss plays a significant role in extracting features, and the gap between semi-hard negative method and random method is tiny in the situation of selecting large number of people in one batch. We also observe that as expected, training with more people helps improve the performance.

We believe this work provides more ideas and inspirations for speaker verification community, and introduces more DNN methods. Although our framework has much room to improve, we think our experimental results will help others understand the task of speaker verification more clearly.

\bibliographystyle{IEEEtran}

\bibliography{mybib}

\begin{thebibliography}{10}
\providecommand{\url}[1]{#1}
\csname url@samestyle\endcsname
\providecommand{\newblock}{\relax}
\providecommand{\bibinfo}[2]{#2}
\providecommand{\BIBentrySTDinterwordspacing}{\spaceskip=0pt\relax}
\providecommand{\BIBentryALTinterwordstretchfactor}{4}
\providecommand{\BIBentryALTinterwordspacing}{\spaceskip=\fontdimen2\font plus
\BIBentryALTinterwordstretchfactor\fontdimen3\font minus
  \fontdimen4\font\relax}
\providecommand{\BIBforeignlanguage}[2]{{%
\expandafter\ifx\csname l@#1\endcsname\relax
\typeout{** WARNING: IEEEtran.bst: No hyphenation pattern has been}%
\typeout{** loaded for the language `#1'. Using the pattern for}%
\typeout{** the default language instead.}%
\else
\language=\csname l@#1\endcsname
\fi
#2}}
\providecommand{\BIBdecl}{\relax}
\BIBdecl

\bibitem{ubm}
D.~Reynolds, T.~F. Quatieri, and R.~B. Dunn, ``Speaker verification using
  adapted gaussian mixture models,'' \emph{Digital Signal Processing}, vol.~10,
  no.~1, pp. 19--41, 2000.

\bibitem{ivector}
N.~Dehak, P.~J. Kenny, R.~Dehak, P.~Dumouchel, and P.~Ouellet, ``Front-end
  factor analysis for speaker verification,'' \emph{IEEE Transactions on Audio,
  Speech, and Language Processing}, vol.~19, no.~4, pp. 788--798, 2011.

\bibitem{softmax}
E.~Variani, X.~Lei, E.~McDermott, I.~L. Moreno, and J.~G. Dominguez, ``Deep
  neural networks for small footprint text-dependent speaker verification,'' in
  \emph{IEEE International Conference on Acoustics, Speech and Signal
  Processing (ICASSP), Florence, Italy}, 2014.

\bibitem{triplet}
C.~Zhang and K.~Koishida, ``End-to-end text-independent speaker verification
  with triplet loss on short utterances,'' in \emph{Interspeech, Stockholm,
  Sweden}, 2017.

\bibitem{plda}
S.~J.~D. Prince and J.~H. Elder, ``Probabilistic linear discriminant analysis
  for inferences about identity,'' in \emph{International Conference on
  Computer Vision (ICCV), Rio de Janeiro, Brazil}, 2007.

\bibitem{facenet}
F.~Schroff, D.~Kalenichenko, and J.~Philbin, ``Probabilistic linear
  discriminant analysis for inferences about identity,'' in \emph{IEEE
  International Conference on Computer Vision and Pattern Recognition (CVPR),
  Boston, MA, USA}, 2015.

\bibitem{improve1}
D.~Cheng, Y.~Gong, S.~Zhou, J.~Wang, and N.~Zheng, ``End-to-end
  text-independent speaker verification with triplet loss on short
  utterances,'' in \emph{IEEE International Conference on Computer Vision and
  Pattern Recognition (CVPR), Las Vegas, Nevada, USA}, 2016.

\bibitem{improve2}
W.~Chen, X.~Chen, J.~Zhang, and K.~Huang, ``Beyond triplet loss: A deep
  quadruplet network for person re-identification,'' in \emph{IEEE
  International Conference on Computer Vision and Pattern Recognition (CVPR),
  Honolulu, Hawaii, USA}, 2017.

\bibitem{improve3}
H.~Alexander, B.~Lucas, and L.~Bastian, ``{In Defense of the Triplet Loss for
  Person Re-Identification},'' \emph{arXiv preprint arXiv:1703.07737}, 2017.

\bibitem{vaegan1}
L.~Tran, X.~Yin, and X.~Liu, ``Representation learning by rotating your
  faces,'' \emph{arXiv preprint arXiv:1705.11136}, 2017.

\bibitem{vaegan2}
A.~Makhzani, N.~J. J.~Shlens, I.~Goodfellow, and B.~Frey, ``{Adversarial
  Autoencoders},'' \emph{arXiv preprint arXiv:1511.05644}, 2015.

\bibitem{embeddings}
D.~Snyder, D.~Garcia-Romero, D.~Povey, and S.~Khudanpur, ``Deep neural network
  embeddings for text-independent speaker verification,'' in \emph{Interspeech,
  Stockholm, Sweden}, 2017.

\bibitem{deepspeaker}
C.~Li*, X.~Ma*, B.~Jiang*, X.~Li*, X.~Zhang, X.~Liu, Y.~Cao, A.~Kannan, and
  Z.~Zhu, ``{Deep Speaker: an End-to-End Neural Speaker Embedding System},''
  \emph{arXiv preprint arXiv:1705.02304}, 2017.

\bibitem{wangdong}
L.~Li, Y.~Chen, Y.~Shi, Z.~Tang, and D.~Wang, ``Deep speaker feature learning
  for text-independent speaker verification,'' in \emph{Interspeech, Stockholm,
  Sweden}, 2017.

\bibitem{original-triplet}
K.~Q. Weinberger and L.~K. Saul, ``Distance metric learning for large margin
  nearest neighbor classification,'' \emph{Journal of Machine Learning
  Research}, vol.~10, pp. 207--244, 2009.

\bibitem{resnet}
K.~He, X.~Zhang, S.~Ren, and J.~Sun, ``Deep residual learning for image
  recognition,'' in \emph{IEEE International Conference on Computer Vision and
  Pattern Recognition (CVPR), Las Vegas, NV, USA}, 2016.

\bibitem{TRISTOUNET}
H.~Bredin, ``Tristounet: Triplet loss for speaker turn embedding,'' in
  \emph{IEEE International Conference on Acoustics, Speech and Signal
  Processing, New Orleans, USA}, 2017.

\bibitem{gan}
I.~Goodfellow, J.~Pouget-Abadie, M.~Mirza, B.~Xu, D.~W. Farley, S.~Ozair,
  A.~Courville, and J.~Bengio, ``{Generative Adversarial Networks},''
  \emph{arXiv preprint arXiv:1406.2661}, 2014.

\bibitem{cgan}
M.~Mirza and S.~Osindero, ``{Conditional Generative Adversarial Nets},''
  \emph{arXiv preprint arXiv:1411.1784}, 2014.

\bibitem{dcgan}
A.~Radford, L.~Metz, and S.~Chintala, ``{Unsupervised Representation Learning
  with Deep Convolutional Generative Adversarial Networks},'' in
  \emph{International Conference on Learning Representation (ICLR), San Juan,
  Puerto Rico}, 2016.

\bibitem{infogan}
X.~Chen, R.~H. Y.~Duan, J.~Schulman, I.~Sutskever, and P.~Abbeel, ``{Infogan:
  Interpretable Representation Learning by Information Maximizing Generative
  Adversarial Nets},'' in \emph{Neural Information Processing Systems (NIPS),
  Barcelona, Spain}, 2016.

\bibitem{wgan}
I.~Gulrajani, F.~Ahmed, M.~Arjovsky, V.~Dumoulin, and A.~Courville, ``{Improved
  Training of Wasserstein GANs},'' \emph{arXiv preprint arXiv:1704.00028},
  2017.

\bibitem{segan}
L.~Yu, W.~Zhang, J.~Wang, and Y.~Yu, ``{SeqGAN: Sequence Generative Adversarial
  Nets with Policy Gradient},'' in \emph{AAAI Conference on Artificial
  Intelligence, San Francisco, California, USA}, 2017.

\bibitem{denoise}
D.~Michelsanti and Z.~Tan, ``Conditional generative adversarial networks for
  speech enhancement and noise-robust speaker verification,'' in
  \emph{Interspeech, Stockholm, Sweden}, 2017.

\bibitem{speech-gan}
A.~Sriram, H.~Jun, Y.~Gaur, and S.~Satheesh, ``{Robust Speech Recognition Using
  Generative Adversarial Networks},'' \emph{arXiv preprint arXiv:1711.01567},
  2017.

\bibitem{tripletgan1}
M.~Zieba and L.~Wang, ``{Training Triplet Networks with GAN},'' \emph{arXiv
  preprint arXiv:1704.02227}, 2017.

\bibitem{tripletgan2}
G.~Cao, Y.~Yang, J.~Lei, C.~Jin, Y.~Liu, and M.~Song, ``{TripletGAN: Training
  Generative Model with TripletLoss},'' \emph{arXiv preprint arXiv:1711.05084},
  2017.

\bibitem{batchnorm}
S.~Ioffe and C.~Szegedy, ``Batch normalization: Accelerating deep network
  training by reducing internal covariate shift,'' in \emph{IEEE International
  Conference on Machine Learning (ICML), Lille, France}, 2015.

\bibitem{sampling-matter}
C.~Wu, R.~Manmatha, A.~J. Smola, and P.~Krähenbühl, ``{Sampling Matters in
  Deep Embedding Learning},'' \emph{arXiv preprint arXiv:1706.07567}, 2017.

\bibitem{librispeech}
V.~Panayotov, G.~Chen, D.~Povey, and S.~Khudanpur, ``Librispeech: An asr corpus
  based on public domain audio books,'' in \emph{IEEE International Conference
  on Acoustics, Speech and Signal Processing (ICASSP), Brisbane, QLD,
  Australia}, 2015.

\bibitem{timit}
J.~S. Garofolo, L.~F. Lamel, W.~M. Fisher, J.~G. Fiscus, D.~S. Pallett, N.~L.
  Dahlgren, and V.~Zue, ``Timit acoustic-phonetic continuous speech corpus,''
  \emph{Linguistic data consortium}, vol.~10, no.~5, 1993.

\end{thebibliography}

\end{document}